# Design optimization of band-pass filter based on parity-time symmetry coupled-resonant

Xinda Lu[1], Nuo Chen[1], Boqing Zhang[1], Haofan Yang[1], Yuntian Chen[1,2], Xinliang Zhang[1,2] and Jing Xu(✉)[1,2]

1 School of Optical and Electronic Information, Huazhong University of Science and Technology, Wuhan 430074, China
2 Wuhan National Laboratory for Optoelectronics, Huazhong University of Science and Technology, Wuhan 430074, China

**Abstract** Integrated optical filter based on microring resonators plays a critical role in many applications, ranging from wavelength division multiplexing and switching to channel routing. Bandwidth tunable filters are capable of meeting the on-demand flexible operations in complex situations, due to their advantages of scalability, multi-function, and energy-saving. It has been investigated recently that parity-time (PT) symmetry coupled-resonant systems can be applied to the bandwidth-tunable filters. However, due to the trade-off between the bandwidth-tunable contrast ratio and insertion loss of system, the bandwidth-tunable contrast ratio of this method is severely limited. Here, the bandwidth-tunable contrast ratio is defined as the maximum bandwidth divided by the minimum bandwidth. In this work, we show that high bandwidth-tunable contrast ratio and low insertion loss of system can be achieved simultaneously by increasing the coupling strength between the input port and the resonant. System characterizations under different coupling states reveal that the low insertion loss can be obtained when the system initially operates at the over-coupling condition. A high bandwidth-tunable contrast ratio PT-symmetry band-pass filter with moderate insertion loss is shown on the Silicon platform. Our scheme provides an effective method to reduce the insertion loss of on-chip tunable filters, which is also applicable to the high-order cascaded microring systems.

## 1 Introduction

Highly integrated photonic technology requires on-demand scalable hardware. A key device for wavelength division multiplexing systems is the functional optical filter with immediate applications for optical communication and switching networks [1,2]. Among the integrated optical devices, microresonator-based instruments have been extensively explored in a broad range of applications, including all-optical signal processing [3–5], lasing [6,7], quantum optics [8], wavelength filtering [9–11], etc. Dynamic tuning of optical microresonators [12–18] has been shown to provide new functionalities for on-chip optical communications and information processing. By simply fabricating microheaters on the microresonators, the resonance of the microresonators can be adjusted through the thermal drift effect [19]. The inherent amplitude-frequency responses of microresonators are usually applied to add, drop, or block selected wavelength channels, which is naturally suitable for designing on-chip bandwidth tunable filters. To date, various tunable filter implementations based on microresonators have been studied, promising to meet the great needs of big data and intelligent optical communications. For instance, employing the Mach-Zehnder interferometers (MZI) combined microresonators for simply tuning the input and output coupling strength to control the resonant linewidth [12]; using multiple microrings cascading structure to reach a large tunable bandwidth [20]; applying tandem microrings for broadening transmittance for band-pass filters [21,22]; and utilizing some other structures to control the tunable-bandwidth [23–28]. All these schemes have analyzed the spectral response of the filter, and the proof-of-principle devices showed the results matched well with the theoretical prediction. However, these designs are relatively complicated.

✉ Corresponding author. E-mail: jing_xu@hust.edu.cn

Parity-time (PT) symmetry [29–31] has shown promise in numerous applications since its first observation in optics [30], such as ring laser [6], sensing [32], optoelectronic oscillator [33], coherent perfect absorption in coupled rings [34], etc. Recent explorations on PT-symmetry in optical coupled-resonant systems open the doors for bandwidth manipulation [34,35]. By introducing appropriate loss on the aimed resonance, i.e., setting the suitable parameters of the coupling strengths between the rings and between the ring and the bus-waveguide, the mode splitting can be well prevented near the exceptional point (EP) where the phase transition occurs [6,31,32,34]. Our recent work has realized a simple and practical bandwidth-tunable band-pass filter by introducing the PT-symmetry coupled-resonant system into the band-pass filters [35]. Benefiting from the flexible adjustability of the system, PT-symmetry enabled band-pass filter has a wide range of applications. However, due to the trade-off between the bandwidth-tunable contrast ratio and insertion loss of the system, PT-symmetry band-pass filter faces a considerable challenge in the high bandwidth-tunable contrast ratio applications. In this work, we reveal the physics behind this trade-off by applying the temporal coupled mode theory (TCMT) [36] and the generalized critical coupling (GCC) condition proposed in [37,38]. The trade-off between the bandwidth-tunable contrast ratio and insertion loss of the system is mitigated by increasing the coupling strength between the input port and the resonant. We further verify the feasibility of our scheme by showing a high bandwidth-tunable contrast ratio PT-symmetry band-pass filter with moderate insertion loss.

## 2 Principle of operation

The PT-symmetry coupled resonant structure with microheaters has been demonstrated to realize the bandwidth tunable band-pass filter [35], the schematic diagram of which is shown in Fig. 1(a). The system consists of a main cavity and an auxiliary cavity with an intercavity coupling rate of $g$, and two bus waveguides coupled to the main cavity and auxiliary cavity, respectively. The signal launches into the waveguide from the bottom left port, and the filtered light leaves from the top right side, as indicated in Fig. 1(a). The intrinsic decay rates of the main ring and the auxiliary ring are $\gamma_{i1}$ and $\gamma_{i2}$, and the coupling decay rates of the main cavity and the auxiliary cavity between the bus-waveguides are $\gamma_{c1}$ and $\gamma_{c2}$, respectively. For coupled resonant device, the sizes of two rings are set to be the same, which results in the same free spectral range ($FSR$), and the same intrinsic decay rate $\gamma_{i1} = \gamma_{i2}$. According to the temporal coupled mode theory

(TCMT) [36], the coupled mode equation of the system can be expressed as $idA/dt = \begin{bmatrix} \omega_1 - i(\gamma_{i1} + \gamma_{c1})/2 & g \\ g & \omega_2 - i(\gamma_{i2} + \gamma_{c2})/2 \end{bmatrix} A - i\sqrt{\gamma_{c1}}(a_{in} \quad 0)^T$, where $A = (a_1 \quad a_2)^T$ is the modal field vector, $a_1$ and $a_2$ are the intracavity fields of the main ring and the auxiliary ring, $a_{in}$ is the input modal field, $\omega_1$ and $\omega_2$ are the resonant frequencies of the main ring and the auxiliary ring, respectively. We define the $\delta$ as the frequency detuning of two rings, satisfying $\delta = (\omega_2 - \omega_1)/2\pi$. The microheater is fabricated upon the auxiliary ring to adjust the $\delta$ by tuning the resonant frequency [19], as shown in the schematic diagram of Fig. 1(b). It should be noted that the initial detuning ($|\delta/FSR| = 0.5$) and the alignment ($\delta/FSR = 0$) of the two rings, corresponding to the positions of minimum and maximum system bandwidth, are represented by the dark and light red lines in Fig. 1(b), respectively. Following our previous work [35], the system is operated close to the EP ($4g \approx \gamma_{c2} - \gamma_{c1}$) at the alignment, where the maximum and minimum bandwidths of the system are $\Delta\omega_{max} = (\gamma_{i1} + \gamma_{c1} + \gamma_{i2} + \gamma_{c2})/2$ and $\Delta\omega_{min} = \gamma_{i1} + \gamma_{c1}$. To evaluate the system performance, the bandwidth-tunable contrast ratio is defined as the system maximum bandwidth divided by minimum bandwidth:

$$\Delta\omega_{ratio} = \frac{\Delta\omega_{max}}{\Delta\omega_{min}} = \frac{1}{2} + \frac{\gamma_{i2} + \gamma_{c2}}{2(\gamma_{i1} + \gamma_{c1})} \tag{1}$$

Equation (1) shows that the system can realize a high bandwidth-tunable contrast ratio when $\gamma_{c2} \gg \gamma_{c1}$. Note that when the intrinsic losses of the rings are small (for high Q factor rings cases), i.e., $\gamma_{c2} \gg \gamma_{i1}$ and $\gamma_{c2} \gg \gamma_{i2}$, the filter can also benefit from a high tunable contrast ratio. However, the Ref [35] only achieves a bandwidth-tunable contrast ratio of about 5.78 (28.8GHz~166.6GHz) with the low system insertion loss, which is limited by the trade-off between bandwidth-tunable contrast ratio and insertion loss of the system.

To ensure the injected signal can be launched into the rings completely, the through port of system is operated in the critical-coupling condition ($\gamma_{i1} \approx \gamma_{c1}$) at the initial detuning [35]. Obviously, when the through port is extinct at the critical coupling, the input signal is entirely transferred to the drop port, resulting in the system insertion loss achieving the minimum. Here, the bandwidth-tunable contrast ratio is found to be

$$\Delta\omega_{ratio} \approx \frac{1}{2} + \frac{\gamma_{c2}}{4\gamma_{c1}} \tag{2}$$

However, due to the influence of auxiliary ring, the coupling state of the through port changes significantly during the detuning. The system insertion loss increases gradually once the coupling state of through port deviates from the critical-coupling condition. According to the GCC condition [37,38], the effective loss rate of the main cavity induced by the auxiliary ring and the output waveguide of auxiliary ring is calculated as $\gamma_{1eff} = 4g^2/(\gamma_{i2} + \gamma_{c2})$, when $\delta = 0$. Note that the effective propagation loss of the main ring is defined as $\gamma_{i1eff} = \gamma_{i1} + \gamma_{1eff}$. Therefore, the coupled mode equation of the dual-ring system can be inferred as

$$\frac{da_1}{dt} = -i\omega_1 a_1 - \frac{\gamma_{i1eff} + \gamma_{c1}}{2} a_1 - \sqrt{\gamma_{c1}} a_{in} \qquad (3)$$

Moreover, considering the influence of the detuning $\delta$ of two rings, the effective propagation loss of the main cavity can be calculated as

$$\gamma_{i1eff} = \gamma_{i1} + \frac{4g^2}{\gamma_{i2} + \gamma_{c2} + \frac{4\delta^2}{\gamma_{i2} + \gamma_{c2}}} \qquad (4)$$

According to the GCC condition, the main cavity is critically coupled to the through-port waveguide when $\gamma_{c1} = \gamma_{i1eff}$. Defining $\Delta\gamma = \gamma_{i1eff}/\gamma_{c1}$ as the degree of deviation between the system coupling state and the critical coupling condition. It is foreseeable that as $\Delta\gamma$ deviates from 0dB, the system coupling state will gradually shift from the critical coupling condition, and the system insertion loss will increase. In other words, the insertion loss of system is positively related to $\Delta\gamma$. And $\Delta\gamma$ reaches the maximum $\Delta\gamma_{max} \approx (\gamma_{c2} - \gamma_{c1})^2/4\gamma_{c2}\gamma_{c1} + 1$, when the two rings are aligned. When the through port is operated in the critical-coupling condition at the initial detuning and the system has a high bandwidth-tunable contrast ratio, the relationship between $\Delta\omega_{ratio}$ and $\Delta\gamma_{max}$ can be obtained according to Eq. (2):

$$\Delta\omega_{ratio} \approx \Delta\gamma_{max} - \frac{1}{2} \qquad (5)$$

Equation (5) indicates that the system bandwidth-tunable contrast ratio $\Delta\omega_{ratio}$ is proportional to $\Delta\gamma_{max}$, confirming that the insertion loss of the system increases with the improving bandwidth-tunable contrast ratios. Furthermore, the evolution of $\Delta\gamma$ as a function of $\delta/FSR$ under different system bandwidth-tunable contrast ratios are plotted with three colored lines in Fig. 1(c). It can be shown that when $|\delta/FSR| = 0.5$, the $\Delta\gamma$ of three lines are all 0dB, which means that the through ports of systems are operated in the critical-coupling condition at the initial detuning. According to the conclusion of the GCC condition, the effective propagation loss $\gamma_{i1eff}$ of main cavity increases as the detuning $\delta$ decreases. The coupling states of the through port will depart from critical-coupling to the under-coupling condition with the reduction of $\delta$, corresponding to the lines in Fig. 1(c), since the effective propagation loss of main cavity gradually increases while the coupling loss of that remains the same. All these changes lead to the insertion loss growing significantly as the coupling state of the through port gradually deviates from the critical-coupling condition. The colored lines drawn in Fig. 1(c) show that the system having a higher bandwidth-tunable contrast ratio corresponds to a larger insertion loss, which matches well with the theoretical expectations (Eq.(5)).

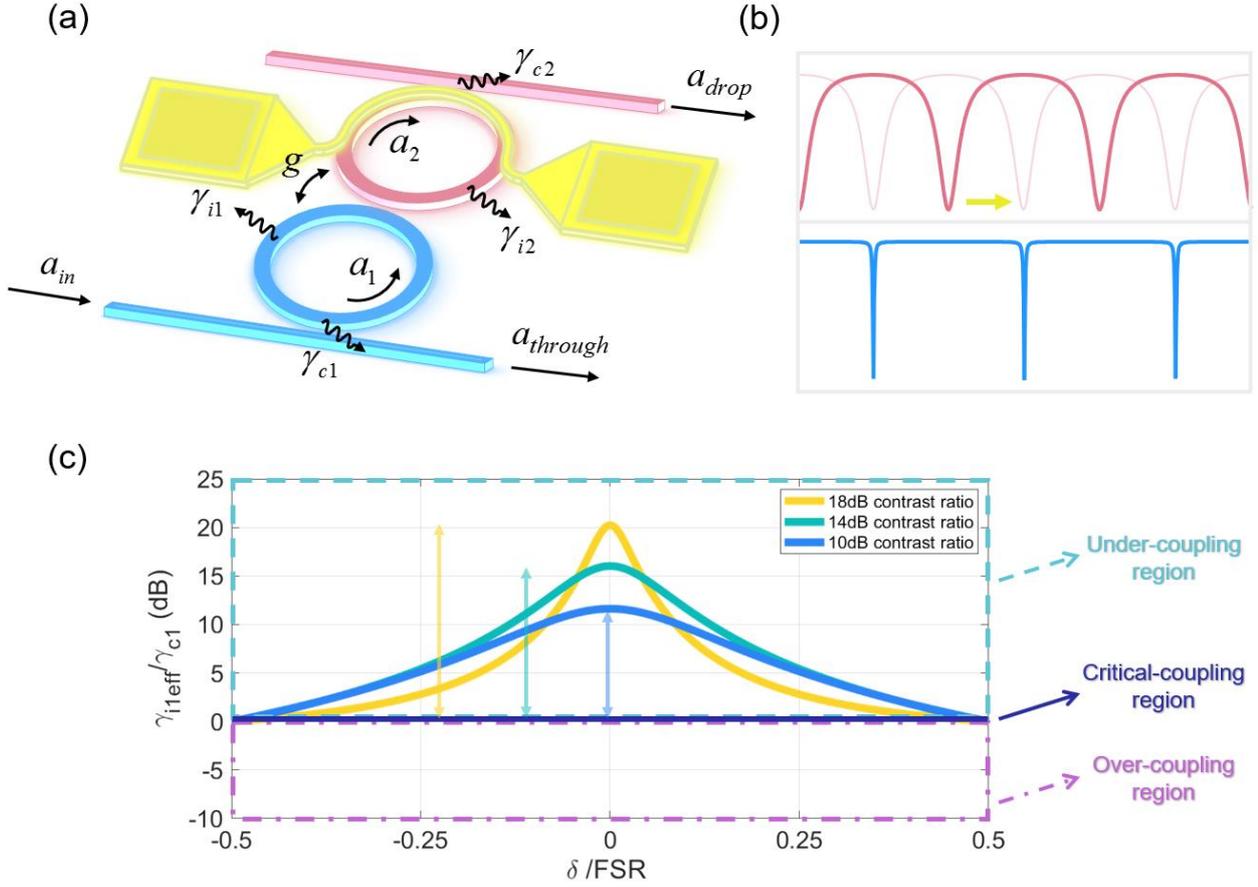

Figure 1. Parity-time (PT) symmetry enabled bandwidth-tunable band-pass filter with the trade-off between the bandwidth tunable contrast ratio and insertion loss. (a) Schematic diagram of the proposed coupled resonant device structure. A microheater (yellow) is fabricated upon the auxiliary ring to tune the resonant frequencies. (b) Schematic of adjusting detuning between two rings. Blue and red lines represent the resonance positions of the main cavity and the auxiliary cavity, respectively. The resonance positions of the auxiliary ring at the initial detuning ($|\delta/FSR| = 0.5$) and the alignment ($\delta/FSR = 0$) are represented by the dark and the light red lines, respectively. (c) Evolution of the ratio of the effective propagation loss to the coupling loss of main cavity ($\gamma_{i1eff}/\gamma_{c1}$) as a function of the detuning $\delta/FSR$. The yellow, green, and blue lines correspond to the systems with bandwidth-tunable contrast ratios of 18dB, 14dB, and 10dB, respectively. The areas circled in the green and purple lines represent the system operating in the under-coupling and over-coupling conditions, respectively. The dark blue line represents that the main cavity is critically coupled to the through port. System working parameters: $\gamma_{i1} = \gamma_{i2} = 2.37 GHz$. Parameters used for the yellow, green, and blue lines are: $\gamma_{c1} = 3.42 GHz, \gamma_{c2} = 369.37 GHz, g = 182.39 GHz$, $\gamma_{c1} = 24.51 GHz, \gamma_{c2} = 1043.7 GHz, g = 505.52 GHz$, and $\gamma_{c1} = 114.23 GHz, \gamma_{c2} = 1847.6 GHz, g = 874.79 GHz$, respectively.

To demonstrate the PT-symmetry coupled-resonant band-pass filter with high bandwidth-tunable contrast ratio, we propose a scheme to mitigate the trade-off between the bandwidth-tunable contrast ratio and insertion loss of the system. Since the deviation of system coupling state from the critical-coupling condition will lead to high insertion loss, the system with low insertion loss can be realized once the coupling state of system can be maintained near the critical-coupling condition throughout the detuning. By increasing the coupling strength between the through port and the main cavity, we modify the evolution of system coupling states with decreasing detuning as slightly over-coupling changes to critical-coupling, finally reaching under-coupling, which keeps the entire insertion loss variation within a small threshold. The schematic diagrams of the systems operating in critical coupling, under-coupling, and over-coupling conditions at the initial detuning are given in Fig. 2(a), (b), (c), respectively. Figure 2(d) indicates the evolution of $\Delta\gamma$ with $\delta/FSR$ under different initial coupling states, where the dark blue, green, and purple lines correspond to the systems with initial coupling states as critical coupling, under-coupling and over-coupling condition, respectively. The dark blue line shows that the insertion loss of the system keeps increasing as $\delta$ decreases. As the detuning decreases, the coupling state of the system, corresponding to the green line in Fig. 2(d), is further away from the critical-coupling condition. Here, the main cavity is strongly under-coupled at the alignment, where the input field is hardly coupled into the system and less field outputs from the drop port, resulting in extremely high system insertion loss. Note that with the decrease of $\delta$, the coupling state of system corresponding to the purple line first approaches critical-coupling and then moves away from the critical-coupling condition, which keeps the entire evolution of the system coupling state near the critical-coupling condition, ensuring that the system insertion loss is maintained in a small threshold. It is foreseeable that the system operating on the purple line can enable the high bandwidth-tunable contrast ratio PT-symmetry coupled-resonant band-pass filter with low insertion loss.

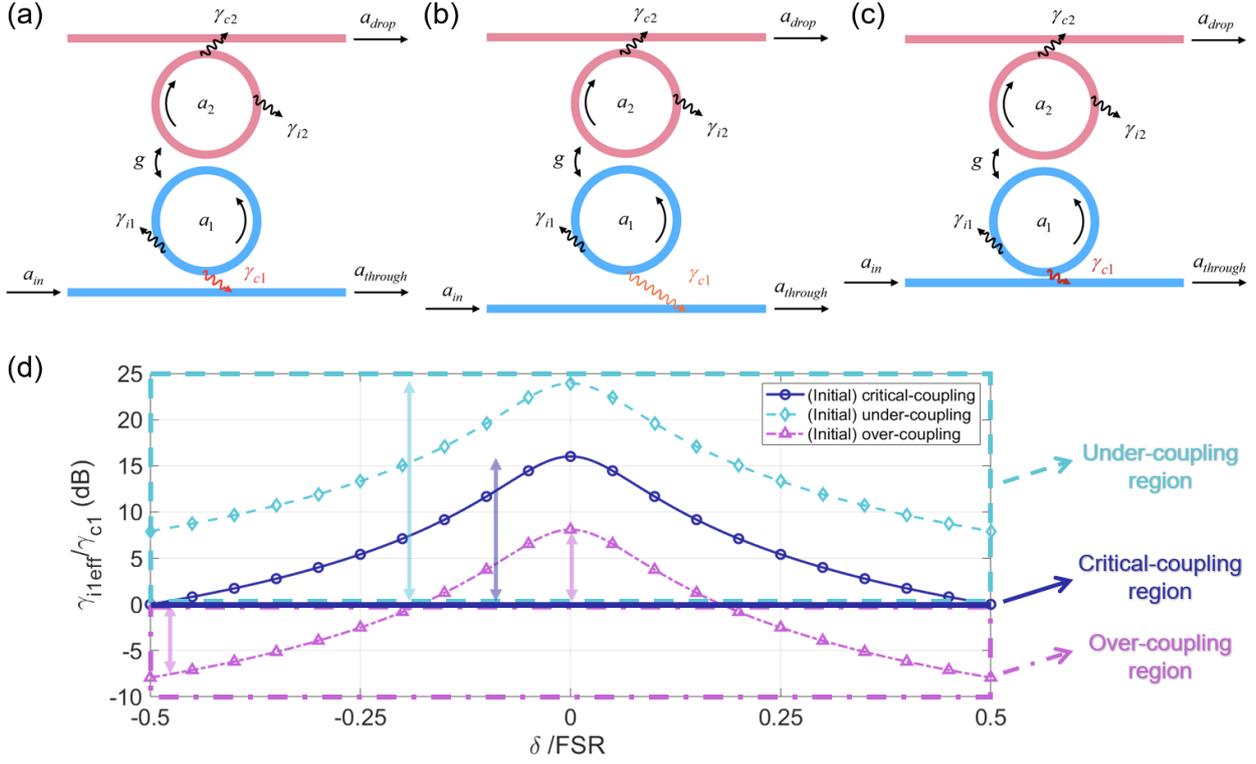

Figure 2. Evolution of system coupling state with detuning under different initial coupling states. (a), (b), and (c) show the schematic diagrams of system operating at the critical-coupling, under-coupling, and over-coupling conditions at initial detuning, respectively. (d) The evolution of the ratio of the effective propagation loss to the coupling loss of main cavity ($\gamma_{i1eff}/\gamma_{c1}$) as a function of the detuning $\delta/FSR$. The dark blue, green, and purple lines correspond to the systems operating at the critical-coupling, under-coupling, and over-coupling conditions at initial detuning, respectively. System working parameters: $\gamma_{i1} = \gamma_{i2} = 2.37 GHz$, $\gamma_{c2} = 1043.7 GHz$, $g = 505.52 GHz$. Parameters used for the dark blue, green, and purple lines are: $\gamma_{c1} = 24.51 GHz$, $\gamma_{c1} = 3.95 GHz$, and $\gamma_{c1} = 151.97 GHz$, respectively.

## 3  Results and Discussion

Although TCMT can characterize the physics behind the system well, it cannot accurately simulate in the entire parameter space. More importantly, it is only suitable for low-loss and weakly coupled systems. In order to evaluate and optimize the system performance under various coupling states, a rigorous analysis of our approach is captured by the transfer matrix method (TMM) [36]. Figure 3(a) shows the notations used in the TMM of the system, where $\sigma_{j=1,2}$ is the roundtrip field attenuation factor of the resonators, $k_2$ is the field coupling coefficient between the resonators, $k_1, k_3$ are the field coupling coefficients of the main cavity and the auxiliary cavity with the waveguides, respectively, $r_{j=1,2,3}$ is the transmission coefficient, where $r_j^2 + k_j^2 = 1$ in the approximation of lossless coupling. $E_{in}$ is the input field, and $E_{through}$ is the output field at the through port, $E_{drop}$ is the output field at the drop port. The transmission spectra of the through port and drop port can be calculated as:

$$T_{through\_TMM} = \left|\frac{E_{through}}{E_{in}}\right|^2 = \left|\frac{r_1 - \sigma_1\tau exp(i\varphi_1)}{1 - r_1\sigma_1\tau exp(i\varphi_1)}\right|^2 \quad (6)$$

$$T_{drop\_TMM} = \left|\frac{E_{drop}}{E_{in}}\right|^2 = \left|\frac{ik_1k_2k_3\sqrt{\sigma_1}exp(i\varphi_1/2)\sqrt{\sigma_2}exp(i\varphi_2/2)}{(1 - r_1\sigma_1\tau exp(i\varphi_1))(1 - r_2\sigma_2 r_3 exp(i\varphi_2))}\right|^2 \quad (7)$$

where $\tau = (r_2 - \sigma_2 r_3 exp(i\varphi_2))/(1 - r_2\sigma_2 r_3 exp(i\varphi_2))$ and $\varphi_{j=1,2}$ is the phase shift of input light traveling per roundtrip in the resonators. The system insertion loss is defined as the maximum loss at the drop port when the resonances are tuned, i.e., the maximal $|T_{drop\_max}|$. It is worth noting that in this work, all the simulations parameters are based on the Silicon platform. Moreover, the lines in Fig. 3(b) indicate the insertion loss of drop port varies with $\delta/FSR$ under the same coupling coefficients $k_2, k_3$ in different initial coupling conditions of critical-coupling, under-coupling and over-coupling, respectively. Note that the initial system coupling states of under-coupling and over-coupling conditions are realized by setting the transmittance of the through port to -6dB at the initial detuning. The solid dark blue line in Fig. 3(b) corresponds to the through port that operates at the critical-coupling condition at initial, showing that the initial insertion loss of drop port is about -1.7dB, where the insertion loss is mainly caused by the intrinsic loss of system. The insertion loss of drop port is slightly reduced with the tuning of $\delta/FSR$, caused by the co-effect of the two supermodes. Note that the effective propagation loss of main cavity is considerably increased at the alignment, where the system represented by the dark blue line enters the strongly under-coupled regime and has a high insertion loss. The green and purple lines feature a larger insertion loss at the initial detuning since the through port is not entirely extinct and some energy escapes. Since the system corresponding to the green line is under the stronger under-coupling condition throughout the tuning process, the overall insertion loss of the green line is larger than that of the dark blue line. Moreover, the purple line indicates that the insertion loss of drop port first decreases to near 0dB and then increases, corresponding to the evolution of the through port coupling states from over-coupling to critical coupling to under-coupling condition, which matches well with theoretical expectations. Figures 4(a) and 4(b) show the transmission spectrum of the drop port in various coupling conditions at the initial detuning and alignment, respectively. It is evident that the system corresponding to the under-coupling condition of through port has the smallest minimum bandwidth, which induces a high bandwidth-tunable contrast ratio. In comparison, the minimum bandwidth of the system in the over-coupling condition is moderately larger, but its overall insertion loss is significantly reduced. By slightly sacrificing the minimum bandwidth of the system, we exploit the simultaneous realization of a high bandwidth-tunable contrast ratio and a low insertion loss, which is of great significance to the PT-symmetry coupled-resonant filter.

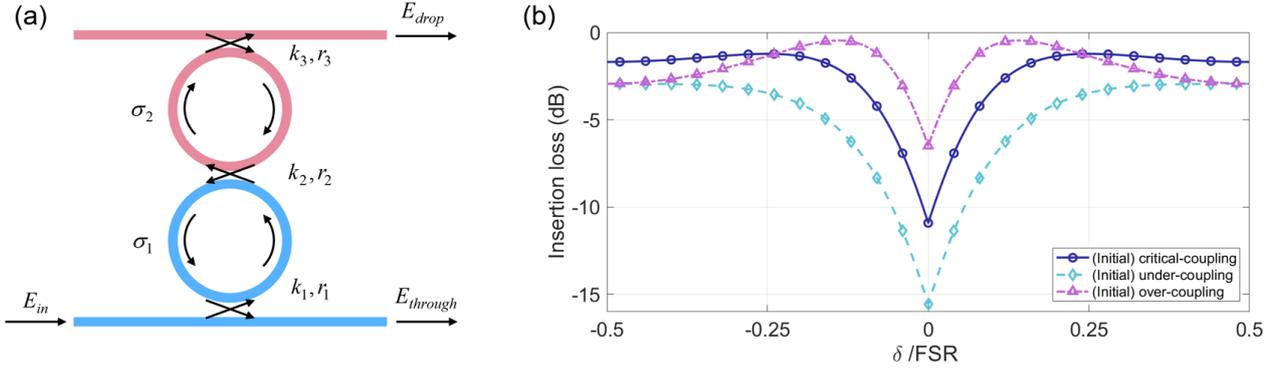

Figure 3. Impact of different coupling states on the system insertion loss. (a) Proposed device structure under the transfer matrix method (TMM). (b) The insertion loss of the drop port as a function of $\delta/FSR$ under different initial coupling conditions. The dark blue, green and purple lines correspond to the systems operating in the critical-coupling, under-coupling, and over-coupling conditions at initial detuning, with the insertion loss range of 1.23~10.91dB, 2.94~15.56dB and 0.46~6.5dB, respectively. Parameters used for the dark blue, green and purple lines are: $k_1 = 0.082, k_2 = 0.206, k_3 = 0.59$ , $k_1 = 0.048, k_2 = 0.206, k_3 = 0.59$ and $k_1 = 0.142, k_2 = 0.206, k_3 = 0.59$, respectively. Parameters of the micro-ring based on Si platform: length of the ring is 62.83um, the intrinsic quality (Q) factor of rings is $5.1 \times 10^5$.

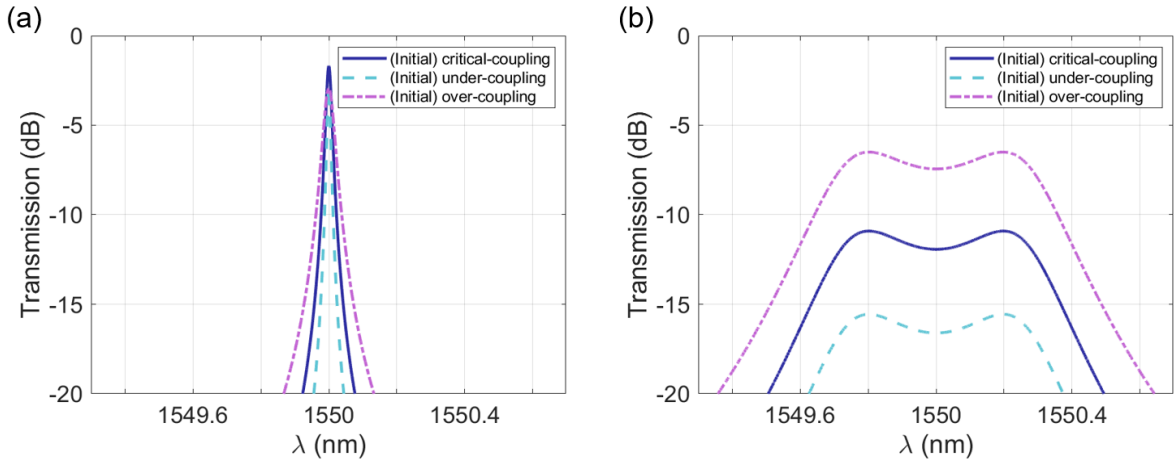

Figure 4. Bandwidth-tunable range of the system under different initial coupling states. (a), (b) The transmission spectra of the drop port at the minimum and maximum bandwidth of the system, where the system bandwidth of different colored lines are 2.35GHz, 1.53GHz, 4.65GHz and 85.02GHz, 84.64GHz, 86.17GHz, respectively. Here, the dark blue, green and purple lines correspond to the systems operating in the critical-coupling, under-coupling, and over-coupling conditions at initial detuning, respectively. Parameters used here are the same as those in Fig. 3.

Furthermore, the system performance under different coupling states is obtained by scanning the coupling coefficients in a substantial parameter space. The diagrams in Fig. 5 show the system bandwidth-tunable contrast ratio as functions of the coupling coefficients $k_2, k_3$, where the contour lines represent the distribution of system insertion loss. Here, the Fig. 5(a), (b), (c) correspond to the bandwidth-tunable contrast ratio evolution of the system operating in the under-coupling, critical-coupling and over-coupling conditions at the initial detuning, respectively. Note that the extinction ratio of the system is above 20dB in a substantial parameter space, which meets the application expectations. The white areas are specifically unsuitable for signal processing, corresponding to the PT-symmetric region, where the excessive splitting in transmission profile (i.e., with more than 3dB in-band ripple compared to the transmission maxima) at the alignment results in signal distortion. It should be noted that each point in the diagrams is obtained by tuning $\delta/FSR$ in the range of -0.5~0.5. In Fig. 5(a), the main cavity is set to the critical coupling condition [37,39] at the initial detuning, where the extinction ratio of the through port is 0. Obviously, the area of the high bandwidth-tunable contrast ratio is scattered in the upper right of Fig. 5(a), which is since the coupling loss of auxiliary cavity is relatively larger than that of main cavity ($k_3 > k_1$), matching well with the predicted result of TCMT. However, the system insertion loss is too high (>11dB) in the high bandwidth-tunable contrast ratio region, significantly reducing system performance. Figures 5(b) and 5(c) show the diagram of system bandwidth-tunable contrast ratio as a function of coupling coefficient at the two initial coupling states of under-coupling and over-coupling, respectively. It can be shown that under the same coupling coefficient $k_2, k_3$, the insertion loss in Fig. 5(c) is smaller than that in Fig. 5(a) and is much smaller than that in Fig. 5(b), which shows the low insertion loss feature of the system operating under the over-coupling condition at initial. It is worth noting that the coupling coefficients $k_2, k_3$ are the same for the three colored points in Fig. 5, corresponding to the three relevant colored lines in Fig. 3(b). The purple dot in Fig. 5(c), operating in the over-coupling condition at the initial detuning, achieves the bandwidth-tunable contrast ratio of ~12.8dB and the system insertion loss of 6.5dB, which indicates that we have realized the high bandwidth-tunable contrast ratio PT-symmetry enabled band-pass filter with moderate insertion loss. It is foreseeable that the PT-symmetry coupled-resonant band-pass filter could be well applied under various platforms, and the system bandwidth-tunable contrast ratio can be improved by reducing the intrinsic loss of materials and increasing the fabrication accuracy. Note that the PT-symmetry enabled band-pass filter can achieve the high bandwidth-tunable contrast ratio with ultra-low insertion loss if the coupling state of through port can be kept at the critical-coupling condition during the detuning, which is expected to be realized by adjusting the coupling loss of main cavity in real time.

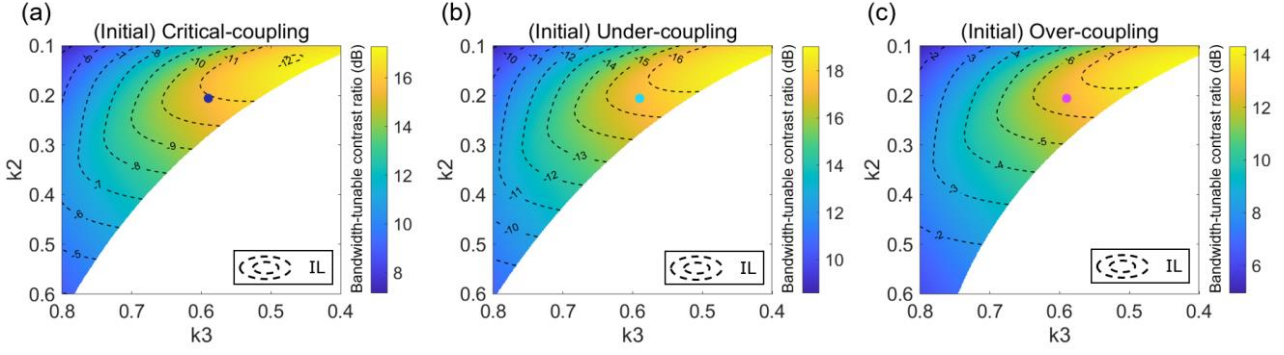

Figure 5. Bandwidth-tunable contrast ratio of the system under different initial coupling states. (a), (b), (c) indicate the evolution of the system bandwidth-tunable contrast ratios as a function of coupling coefficient $k_2, k_3$, corresponding to the systems operating in the critical-coupling, under-coupling, and over-coupling conditions at initial detuning, respectively. The contour lines represent the distribution of system insertion loss (*IL*). The dark blue, green, and purple points represent the systems, corresponding to the same colored lines in Fig. 3 and Fig. 4, with the insertion loss of 10.91dB, 15.56dB, and 6.5dB, respectively. Parameters used here are the same as those in Fig. 3.

# 4 Conclusions

In conclusion, we theoretically demonstrate a scheme that mitigates the trade-off between the bandwidth-tunable contrast ratio and insertion loss of the PT-symmetry enabled band-pass filter. A high bandwidth-tunable contrast ratio system with moderate insertion loss is obtained simultaneously. We reveal the trade-off in PT-symmetry coupled-resonant band-pass filter systems through TCMT and GCC condition. Moreover, the effect of the through port coupling states at initial detuning on the system performance is discussed. The trade-off between the bandwidth-tunable contrast ratio and insertion loss of the system is solved by increasing the coupling strength between the through port and the main cavity, which is useful to realize the band-pass filter with high bandwidth-tunable contrast ratio. It is worth noting that we realize a reduction (-4.4dB) in the insertion loss by slightly sacrificing the minimum bandwidth (2.3GHz) of system, which is proved by TMM. The low insertion loss features of the system initially operating in the over-coupled condition are demonstrated by scanning the coupling coefficients in a substantial parameter space. Note that the system insertion loss can be further reduced if the coupling loss of main cavity could be adjusted in real time. The implementation of our scheme is simple and elegant, and only the thermal tuning is sufficient. The system applies to various material platforms for experimental implementations, showing potential in a wide range of applications. Finally, the scheme we proposed solves the core problem of PT-symmetry enabled band-pass filter, making it applicable to more scenarios.

**REFERENCE**

[1] L.-W. Luo, N. Ophir, C. P. Chen, L. H. Gabrielli, C. B. Poitras, K. Bergmen, and M. Lipson, *WDM-Compatible Mode-Division Multiplexing on a Silicon Chip*, Nat. Commun. **5**, 3069 (2014).

[2] S. Xiao, M. H. Khan, H. Shen, and M. Qi, *Multiple-Channel Silicon Micro-Resonator Based Filters for WDM Applications*, Opt. Express **15**, 7489 (2007).

[3] S. Wabnitz and B. J. Eggleton, editors , *All-Optical Signal Processing*, Vol. 194 (Springer International Publishing, Cham, 2015).

[4] A. E. Willner, S. Khaleghi, M. R. Chitgarha, and O. F. Yilmaz, *All-Optical Signal Processing*, J. Light. Technol. **32**, 660 (2014).

[5] F. Morichetti, A. Canciamilla, C. Ferrari, A. Samarelli, M. Sorel, and A. Melloni, *Travelling-Wave Resonant Four-Wave Mixing Breaks the Limits of Cavity-Enhanced All-Optical Wavelength Conversion*, Nat. Commun. **2**, 296 (2011).

[6] B. Peng, Ş. K. Özdemir, S. Rotter, H. Yilmaz, M. Liertzer, F. Monifi, C. M. Bender, F. Nori, and L. Yang, *Loss-Induced Suppression and Revival of Lasing*, Science **346**, 328 (2014).

[7] L. Feng, Z. J. Wong, R.-M. Ma, Y. Wang, and X. Zhang, *Single-Mode Laser by Parity-Time Symmetry Breaking*, Science **346**, 972 (2014).

[8] Y. Liu, C. Wu, X. Gu, Y. Kong, X. Yu, R. Ge, X. Cai, X. Qiang, J. Wu, X. Yang, and P. Xu, *High-Spectral-Purity Photon Generation from a Dual-Interferometer-Coupled Silicon Microring*, Opt. Lett. **45**, 73 (2020).

[9] D. Liu, H. Xu, Y. Tan, Y. Shi, and D. Dai, *Silicon Photonic Filters*, Microw. Opt. Technol. Lett. **63**, 2252 (2021).

[10] M. A. Popovíc, T. Barwicz, M. R. Watts, P. T. Rakich, L. Socci, E. P. Ippen, F. X. Kärtner, and H. I. Smith, *Multistage High-Order Microring-Resonator Add-Drop Filters*, Opt. Lett. **31**, 2571 (2006).

[11] B. E. Little, S. T. Chu, P. P. Absil, J. V. Hryniewicz, F. G. Johnson, F. Seiferth, D. Gill, V. Van, O. King, and M. Trakalo, *Very High-Order Microring Resonator Filters for WDM Applications*, IEEE Photonics Technol. Lett. **16**, 2263 (2004).

[12] L. Chen, N. Sherwood-Droz, and M. Lipson, *Compact Bandwidth-Tunable Microring Resonators*, 3 (n.d.).

[13] Y. Zhang, Q. Liu, C. Mei, D. Zeng, Q. Huang, and X. Zhang, *Proposal and Demonstration of a Controllable Q Factor in Directly Coupled Microring Resonators for Optical Buffering Applications*, Photonics Res. **9**, 2006 (2021).

[14] T. Tanabe, M. Notomi, H. Taniyama, and E. Kuramochi, *Dynamic Release of Trapped Light from an Ultrahigh- Q Nanocavity via Adiabatic Frequency Tuning*, Phys. Rev. Lett. **102**, 043907 (2009).

[15] V. M. Menon, W. Tong, and S. R. Forrest, *Control of Quality Factor and Critical Coupling in Microring Resonators Through Integration of a Semiconductor Optical Amplifier*, IEEE Photonics Technol. Lett. **16**, 1343 (2004).

[16] M. J. Strain, C. Lacava, L. Meriggi, I. Cristiani, and M. Sorel, *Tunable Q-Factor Silicon Microring Resonators for Ultra-Low Power Parametric Processes*, Opt. Lett. **40**, 1274 (2015).

[17] Y. Tanaka, J. Upham, T. Nagashima, T. Sugiya, T. Asano, and S. Noda, *Dynamic Control of the Q Factor in a Photonic Crystal Nanocavity*, Nat. Mater. **6**, 862 (2007).

[18] S. Manipatruni, C. B. Poitras, Q. Xu, and M. Lipson, *High-Speed Electro-Optic Control of the Optical Quality Factor of a Silicon Microcavity*, Opt. Lett. **33**, 1644 (2008).

[19] P. Dong, N.-N. Feng, D. Feng, W. Qian, H. Liang, D. C. Lee, B. J. Luff, T. Banwell, A. Agarwal, P. Toliver, R. Menendez, T. K. Woodward, and M. Asghari, *GHz-Bandwidth Optical Filters Based on High-Order Silicon Ring Resonators*, Opt. Express **18**, 23784 (2010).

[20] T. Dai, A. Shen, G. Wang, Y. Wang, Y. Li, X. Jiang, and J. Yang, *Bandwidth and Wavelength Tunable Optical Passband Filter Based on Silicon Multiple Microring Resonators*, Opt. Lett. **41**, 4807 (2016).

[21] C. L. Manganelli, P. Pintus, F. Gambini, D. Fowler, M. Fournier, S. Faralli, C. Kopp, and C. J. Oton, *Large-FSR*

*Thermally Tunable Double-Ring Filters for WDM Applications in Silicon Photonics*, IEEE Photonics J. **9**, 1 (2017).

[22] J. R. Ong, R. Kumar, and S. Mookherjea, *Ultra-High-Contrast and Tunable-Bandwidth Filter Using Cascaded High-Order Silicon Microring Filters*, IEEE Photonics Technol. Lett. **25**, 1543 (2013).

[23] Y. Ding, M. Pu, L. Liu, J. Xu, C. Peucheret, X. Zhang, D. Huang, and H. Ou, *Bandwidth and Wavelength-Tunable Optical Bandpass Filter Based on Silicon Microring-MZI Structure*, Opt. Express **19**, 6462 (2011).

[24] G. Poulopoulos, G. Giannoulis, N. Iliadis, D. Kalavrouziotis, D. Apostolopoulos, and H. Avramopoulos, *Flexible Filtering Element on SOI With Wide Bandwidth Tunability and Full FSR Tuning*, J. Light. Technol. **37**, 300 (2019).

[25] P. Orlandi, F. Morichetti, M. J. Strain, M. Sorel, P. Bassi, and A. Melloni, *Photonic Integrated Filter With Widely Tunable Bandwidth*, J. Light. Technol. **32**, 897 (2014).

[26] H. Wang, J. Dai, H. Jia, S. Shao, X. Fu, L. Zhang, and L. Yang, *Polarization-Independent Tunable Optical Filter with Variable Bandwidth Based on Silicon-on-Insulator Waveguides*, Nanophotonics **7**, 1469 (2018).

[27] H. L. R. Lira, C. B. Poitras, and M. Lipson, *CMOS Compatible Reconfigurable Filter for High Bandwidth Non-Blocking Operation*, Opt. Express **19**, 20115 (2011).

[28] J. Yao and M. C. Wu, *Bandwidth-Tunable Add–Drop Filters Based on Micro-Electro-Mechanical-System Actuated Silicon Microtoroidal Resonators*, Opt. Lett. **34**, 2557 (2009).

[29] R. El-Ganainy, K. G. Makris, M. Khajavikhan, Z. H. Musslimani, S. Rotter, and D. N. Christodoulides, *Non-Hermitian Physics and PT Symmetry*, Nat. Phys. **14**, 11 (2018).

[30] C. E. Rüter, K. G. Makris, R. El-Ganainy, D. N. Christodoulides, M. Segev, and D. Kip, *Observation of Parity–Time Symmetry in Optics*, Nat. Phys. **6**, 192 (2010).

[31] B. Peng, Ş. K. Özdemir, F. Lei, F. Monifi, M. Gianfreda, G. L. Long, S. Fan, F. Nori, C. M. Bender, and L. Yang, *Parity–Time-Symmetric Whispering-Gallery Microcavities*, Nat. Phys. **10**, 394 (2014).

[32] W. Chen, Ş. Kaya Özdemir, G. Zhao, J. Wiersig, and L. Yang, *Exceptional Points Enhance Sensing in an Optical Microcavity*, Nature **548**, 192 (2017).

[33] J. Zhang and J. Yao, *Parity-Time–Symmetric Optoelectronic Oscillator*, Sci. Adv. **4**, eaar6782 (2018).

[34] C. Wang, W. R. Sweeney, A. D. Stone, and L. Yang, *Coherent Perfect Absorption at an Exceptional Point*, Science **373**, 1261 (2021).

[35] B. Zhang, N. Chen, X. Lu, Y. Hu, Z. Yang, X. Zhang, and J. Xu, *Bandwidth Tunable Optical Bandpass Filter Based on Parity-Time Symmetry*, Micromachines **13**, 89 (2022).

[36] V. Van, *Optical Microring Resonators: Theory, Techniques, and Applications*, 0 ed. (CRC Press, Boca Raton, FL : CRC Press, Taylor & Francis Group, [2017] |, 2016).

[37] Y. Hu, M. Yu, D. Zhu, N. Sinclair, A. Shams-Ansari, L. Shao, J. Holzgrafe, E. Puma, M. Zhang, and M. Lončar, *On-Chip Electro-Optic Frequency Shifters and Beam Splitters*, Nature **599**, 587 (2021).

[38] M. Lončar, Y. Hu, M. Yu, B. Buscaino, N. Sinclair, D. Zhu, R. Cheng, A. Shams-Ansari, L. Shao, M. Zhang, and J. Kahn, Integrated High-Efficiency and Broadband Electro-Optic Frequency Comb Generators, preprint, In Review, 2022.

[39] K. Komagata, A. Tusnin, J. Riemensberger, M. Churaev, H. Guo, A. Tikan, and T. J. Kippenberg, *Dissipative Kerr Solitons in a Photonic Dimer on Both Sides of Exceptional Point*, Commun. Phys. **4**, 159 (2021).